\documentclass[prd,eqsecnum,showpacs,showkeys,floatfix,
               preprint,tightenlines]{revtex4}

\usepackage{amsmath}
\usepackage{amsfonts}
\usepackage{latexsym}
\usepackage{graphicx}
\usepackage{mathrsfs}

\newcommand{\version}{v5}           
\newcommand{\no}{\nonumber}
\newcommand{\bR}{\mathbb{R}}

\newcommand{\bZ}{\mathbb{Z}}
\newcommand{\cS}{\mathscr{S}}

\def\e{{\rm e}}

\renewcommand{\i}{{\rm i}}
\newcommand{\ri}{{\rm i}}
\newcommand{\tfr}[2]{{\textstyle \frac{#1}{#2}}}

\newcommand{\dx}{\!{\rm d}^4x\,\,}
\newcommand{\dvx}{\!{\rm d}^3 \vec x\,\,}
\newcommand{\dvy}{\!{\rm d}^3 \vec y\,\,}
\newcommand{\dint}[1]{\!{\rm d}#1\,\,}

\newcommand{\tr}{\text{tr}}

\newcommand{\vect}    [1]{\left( \begin{array}{c} #1 \end{array} \right)}

\newcommand{\twomat}  [1]{\left( \begin{array}{cc} #1 \end{array} \right)}

\newcommand{\al}{\alpha}
\newcommand{\da}{ {\dot \alpha} }
\newcommand{\pr}{\partial}

\newcommand{\Ifermion}{I_{\rm \,fermion}}
\newcommand{\id}{\mathbf{1}}

\newcommand {\rhs} {right-hand side}

\newcommand {\LC}  {Levi--Civita}
\newcommand {\KG}  {Klein--Gordon}
\newcommand {\CS}  {Chern--Simons}
\newcommand{\YM}   {Yang--Mills}

\newcommand{\SM}   {Standard Model}

\begin{document}
\noindent Phys. Rev. D {\bf 70}, 045020 (2004) 
\hspace*{\fill}  hep-th/0312032, KA--TP--10--2003 (\version)
\vspace*{2\baselineskip}
\title{Spacetime foam, CPT anomaly, and photon propagation}

\author{Frans R.\ Klinkhamer}
\email{frans.klinkhamer@physik.uni-karlsruhe.de}
\author{Christian Rupp}
\email{cr@particle.uni-karlsruhe.de}
\affiliation{Institut f\"ur Theoretische Physik,\\
             Universit\"at Karlsruhe (TH),\\
             76128 Karlsruhe, Germany}

\begin{abstract}
The CPT anomaly of certain chiral gauge theories 
has been established previously 
for flat multiply connected spacetime manifolds 
$M$ of the type $\bR^3 \times S^1$, 
where the noncontractible loops have a minimal length. 
In this article, we show that 
the CPT anomaly also occurs for manifolds where the 
noncontractible loops can be arbitrarily small. Our basic 
calculation is performed for a flat 
noncompact manifold with a single ``puncture,'' namely 
$M=\bR^2 \times (\bR^2 \setminus \{0\})$. 
A hypothetical spacetime foam might have many such punctures
(or other structures with similar effects).
Assuming the multiply connected structure of the foam to be 
time independent, we present a simple model for photon propagation, 
which generalizes the single-puncture result.
This model leads to a modified dispersion law of the photon. 
Observations of 
high-energy photons (gamma-rays) from explosive extragalactic events
can then be used to place an upper bound on the typical length scale of
these punctures. 
\end{abstract}

\pacs{11.15.-q, 04.20.Gz, 11.30.Cp, 98.70.Rz}
\keywords{Gauge field theories, Spacetime topology, 
          Lorentz and CPT noninvariance, Gamma-ray sources}

\maketitle

\section{Introduction}
\label{sec:intro}

Chiral gauge theories may display an anomalous breaking of Lorentz and
CPT invariance if these theories are defined on a spacetime manifold
$M$ with nontrivial topology \cite{K00, KS02}. 
One example of an appropriate manifold is $M=\bR^3\times S^1$, 
as long as the chiral fermions have the correct boundary conditions
(spin structure) over the compact space dimension.
This so-called CPT anomaly has been established also for
two-dimensional ``chiral'' $U(1)$ gauge theories over the torus $T^2$, where
the Euclidean effective action is known exactly \cite{KN01, KM01}.
The main points of the CPT anomaly have been reviewed in Ref.\
\cite{K01wigner} and some background material can be found 
in Ref.\ \cite{KR03}.

The crucial ingredient of the CPT anomaly is the existence of 
a compact, separable space dimension with appropriate spin structure. 
For the case that this dimension is closed, there 
are corresponding noncontractible loops
over the spacetime manifold, which must therefore be multiply connected.
Up till now, attention has been focused on flat spacetime manifolds $M$
with topology $\bR^3\times S^1$ or $\bR\times T^3$ and with 
Minkowski metric $g_{\mu\nu}(x)=\eta_{\mu\nu}$ and trivial vierbeins
$e_\mu^a(x)=\delta_\mu^a\,$. 
Here, the noncontractible loops occur at the very largest scale.
But there could also be noncontractible loops from nontrivial
topology at the very smallest scale, possibly related to the so-called
spacetime foam \cite{W57, W68, H78, HPP80, Fetal90, H91, GH93, V96}.
The main question is then whether or not a foam-like structure of
spacetime could give rise to some kind of CPT anomaly.

In the present article we give a positive answer to this question. That is, 
we establish the CPT anomaly for one of the simplest possible manifolds of 
this type, namely $M=\bR\times M_3$, where the three-dimensional space manifold 
$M_3$ is flat Euclidean space $\bR^3$ with one straight line $\bR$ removed.
This particular manifold $M_3$ has a single ``puncture'' and 
arbitrarily small noncontractible loops.

But it is, of course, an open question whether or not spacetime really has a
foam-like structure and, if so, with which characteristics.
Let us just consider one possibility. Suppose that the spacetime manifold  
$\mathcal{M}$ is Lorentzian, $\mathcal{M}=\bR\times \mathcal{M}_3$, and that 
the topology of $\mathcal{M}_3$ is multiply connected and time-independent 
(here, time corresponds to the coordinate $x^0\in \bR$). 
The idea is that the multiply connectedness of $\mathcal{M}_3$ is ``hard-wired.''  
The advantage of this restriction is that, for the moment, 
we do not have to deal with the contentious issue of topology change; 
cf.\ Refs.\ \cite{H91,V96}.

Physically, we are then interested in the long-range effects (via the CPT
anomaly) of this foam-like structure of space on the propagation of
light.  The exact calculation of these anomalous effects is, however, 
not feasible for a spacetime manifold $\mathcal{M}$ with many punctures
(or similar structures).  
We, therefore, introduce a model for the photon field over $\bR^4$
which incorporates the basic
features found for the case of a single puncture. The model is
relatively simple and we can study the issue of photon propagation
in the large-wavelength limit.

The outline of this article is as follows. In Sec.\ \ref{sec:SingleDefect}, 
we establish the CPT anomaly for a flat spacetime manifold, where 
the spatial hypersurfaces have a single puncture
corresponding to a static linear ``defect.''
(A similar result is obtained for a space manifold 
with a single wormhole, which corresponds to a static point-like defect.
The details of this calculation are relegated to 
Appendix \ref{sec:single_wormhole}.)

In Sec.\ \ref{sec:Models}, we present a model for the photon field which
generalizes the result of a single puncture (or wormhole).
We also give a corresponding model for a real scalar field.
Both models involve a ``random'' 
background field over $\bR^4$, denoted by $g_0 (x)$ for 
the scalar model and  by $g_1 (x)$ for the photon model. 
For the photon model in
particular, the random background field $g_1 (\vec x)$ is believed to
represent the effects of a static spacetime foam
(a more or less realistic example for the case of permanent wormholes is
given in Appendix \ref{sec:wormhole_gas}). In Sec.\ \ref{sec:RandomBackground}, 
we discuss some assumed properties of these random background fields.

In Sec. \ref{sec:Dispersion}, we calculate the dispersion laws for 
the scalar and photon models of Sec.\ \ref{sec:Models}.
In Sec.\ \ref{sec:explimit}, we use observations of gamma-rays
from explosive extragalactic events
to place an upper bound on the typical length scale of the random
background field $g_1 (\vec x)$ and, thereby, on the typical length
scale of the postulated foam-like structure.
In Sec.\ \ref{sec:conclusion}, we present some concluding remarks.

\begin{widetext} 
\section{CPT anomaly from a single defect of space} 
\label{sec:SingleDefect}

\subsection{Chiral gauge theory and punctured manifold}
\label{sec:SPtheory}

In this section, three-space is taken to be the punctured manifold 
$M_3=\bR \times (\bR^2 \setminus \{0\}) = \bR^3 \setminus \bR\,$. 
The considered three-space may be said to have a linear ``defect,'' 
just as a type-II superconductor can have a single vortex line
(magnetic flux tube).  The corresponding four-dimensional spacetime 
manifold $M=\bR \times M_3$ is orientable and has Minkowski metric
$(\eta_{\mu\nu})\equiv{\rm diag}(1,-1,-1,-1)$ and trivial vierbeins
$e_\mu^a(x)=\delta_\mu^a\,$.  This particular spacetime
manifold $M$ is, of course, geodesically incomplete, 
but the affected geodesics constitute a set of measure zero. At the
end of Sec.\ \ref{sec:SPanomaly}, we briefly discuss another manifold,
$N$, which is both multiply connected and geodesically complete.

The main interest here is in quantized fermion fields 
which propagate over the predetermined spacetime manifold
$M$ and which are coupled to a given classical gauge field.
We will use cylindrical coordinates around the removed line in three-space and 
will rewrite the four-dimensional  
field theory as a three-dimensional one with infinitely many 
fermionic fields, at least for an appropriate choice of gauge field.
This procedure is analogous to the one used for the original CPT anomaly
from the multiply connected manifold $\bR^3 \times S^1$.

The gauge field is written as $B_{\mu}(x) = g\, B_\mu^{b}(x)\,T^b$,
with an implicit sum over $b$, where $g$ is  the gauge coupling constant and 
the $T^b$ are the anti-Hermitian generators of 
the Lie algebra, normalized by 
${\rm tr}\, (T^b T^c)$ $=$ $- \, \frac{1}{2}\, \delta^{bc}$.   
For the matter fields, we take a single
complex multiplet of left-handed Weyl fermions $\psi_\al (x)$.
As a concrete case, we consider the $SO(10)$ gauge theory with 
left-handed Weyl fermions in the {\bf 16} representation. 
This particular chiral gauge theory
includes, of course, the \SM~with one family of quarks and leptons. 
Incidentally, the anomalous effects of this section do not occur for
vectorlike theories such as Quantum Electrodynamics.

In short, the theory considered has 
\begin{equation}
\left[\,  G\,,\, R_L\,,\, M\,,\, e_\mu^a(x)\, \right] = 
\left[\, SO(10)\,,\,\mathbf{16}\,,\,\bR^4\setminus \bR^2\,,\,\delta^a_\mu\, 
\right]\,,
\end{equation}
where $G$ denotes the gauge group, $R_L$ the representation of the
left-handed Weyl fermions, $M$ the spacetime manifold, and
$e_\mu^a(x)$ the vierbeins at spacetime point $x \in M$.
The action for the fermionic fields reads
\begin{equation}
\Ifermion = \int_{M} \dx \bar \psi_\da \; \delta_a^\mu \;
\ri\, \bar \sigma^{a\da\al}\, \left(\pr_\mu + B_\mu\right)\,\psi_\al\,,
\label{Gfermion}
\end{equation}
with
\begin{equation}
(\bar\sigma^{a\da\al}) = \left(\, \id, -\vec \sigma \,\right)\,,
\end{equation}
in terms of the Pauli spin matrices $\sigma^1$, $\sigma^2$, $\sigma^3$. 
Natural units with $\hbar$ $=$ $c$ $=$ $1$ are used throughout, 
except when stated otherwise. 

Next, introduce cylindrical coordinates 
$(\rho,\phi, z\equiv x^3, t \equiv x^0)$, with
\begin{align} \label{cylindricalcoord}
x^1 &\equiv \rho \cos \phi\,, & x^2 &\equiv \rho \sin\phi \,,\nonumber\\
\partial_\rho &\equiv \cos\phi       \; \pr_1 + \sin\phi      \; \pr_2 \,,&
\partial_\phi &\equiv -\rho \sin\phi \; \pr_1 + \rho \cos\phi \; \pr_2 \,,\nonumber\\
\bar \sigma^\rho &\equiv \cos\phi \; \bar \sigma^1 + \sin\phi\; \bar \sigma^2\,, &
\bar \sigma^\phi &\equiv -\sin\phi\; \bar \sigma^1 + \cos\phi\; \bar \sigma^2\,.
\end{align}
The two spin matrices of Eq.~(\ref{cylindricalcoord}) and $\bar\sigma^z$ 
are explicitly
\begin{equation}
(\bar \sigma^{\rho\da\al}) = \twomat{0 & -\e^{-\ri\phi}\\ -\e^{\ri\phi}
  &0},\;\;
(\bar\sigma^{\phi\da\al}) = \twomat{0 & \ri\, \e^{-\ri\phi}\\
  -\ri\, \e^{\ri\phi} & 0},\;
(\bar \sigma^{z\da\al}) = \twomat{-1 & 0\\0  &+1}.
\end{equation}

For the spinorial wave functions in terms of cylindrical coordinates,
we take the antiperiodic boundary condition
\begin{equation}
\psi(\rho, \phi+2\pi, z, t) = -\psi(\rho, \phi, z, t)\,.
\end{equation}
An appropriate \emph{Ansatz} for the fermion fields is then
\begin{subequations}
\begin{align}
\psi_\al(\rho, \phi, z, t) &= \sum_{n=-\infty}^\infty \Bigl\{
  \e^{+\ri (n-1/2) \phi} \,  \chi_n^{(+)}(\rho, z, t) \,\psi_\al^{(+)}  \,
 \,+ \, \e^{+\ri (n+1/2) \phi}\, \chi_n^{(-)}(\rho, z, t) \,\psi_\al^{(-)}
   \Bigr\}\,, \\ 
\bar \psi_\da(\rho, \phi, z, t) &= \sum_{n=-\infty}^\infty \Bigl\{
  \e^{-\ri (n-1/2) \phi} \, \bar \chi_n^{(+)}(\rho, z, t) \,\bar \psi_\da^{(+)}  \,
 \,+ \, \e^{-\ri (n+1/2) \phi}\,\bar  \chi_n^{(-)}(\rho, z, t)\,\bar \psi_\da^{(-)}
  \Bigr\}\,,
\end{align}\label{Ansatz}
\end{subequations}
with constant spinors
\begin{equation}
\bigl(\psi^{(+)}_\al\bigr) = \vect{1\\0}\,,\quad 
\bigl(\psi^{(-)}_\al\bigr)=\vect{0\\1}\,,\quad
\bigl(\bar \psi^{(+)}_\da\bigr) = \vect{1\\0}\,,\quad 
\bigl(\bar \psi^{(-)}_\da\bigr) =\vect{0\\1}\,,
\end{equation}
and anticommuting fields $\chi^{(\pm)}_n$ which depend only on 
the coordinates $\rho$, $z$, and $t$. 
The unrestricted fields 
$ \chi_0^{(\pm)}(\rho, z, t)$ and $\bar \chi_0^{(\pm)}(\rho, z, t)$
of Eqs.~(\ref{Ansatz}a,b) will play an important role in 
the next subsection.

\subsection{CPT anomaly}   
\label{sec:SPanomaly}

In order to demonstrate the existence of the CPT anomaly, it suffices
to consider a special class of gauge fields (denoted by a prime)
which are $\phi$-independent and have vanishing 
components in the $\phi$ direction; cf. Ref. \cite{K00}.
Specifically, we consider in this subsection the following gauge
fields:
\begin{equation}
B^\prime_\phi = B^\prime_\phi (\rho,z,t) =0\,,\quad
B^\prime_m    = B^\prime_m (\rho,z,t)\,,\quad m= \rho,z,t. 
\label{Bprime}
\end{equation}
Using the \emph{Ansatz} (\ref{Ansatz}) and integrating 
over the azimuthal angle $\phi$, the action (\ref{Gfermion}) 
can be written as
\begin{align}       
\Ifermion = &2\pi\ri \sum_{n=-\infty}^{+\infty}\, 
\int_\epsilon^\infty  \!{\rm d}\rho
\, \rho \, \int^\infty_{-\infty}  \!{\rm d}z\, 
\int^\infty_{-\infty} \!{\rm d} t\, \, \Bigl\{ 
\bar \chi_n^{(+)} \bigl(\,\partial_t + B^\prime_t\,\bigr) \chi_n^{(+)} + 
\bar \chi_n^{(-)}
\bigl(\,\partial_t+B^\prime_t\,\bigr) \chi_n^{(-)} \nonumber \\
&
-\bar \chi_n^{(+)} \bigl(\,\partial_\rho + B^\prime_\rho\,\bigr) \chi_n^{(-)} - \bar
\chi_n^{(-)} \bigl(\,\partial_\rho + B^\prime_\rho\,\bigr) \chi_n^{(+)} 
-\bar \chi_n^{(+)} \bigl(\,\partial_z+B^\prime_z\,\bigr) \chi_n^{(+)} \nonumber\\[2mm]
&
+ \bar \chi_n^{(-)} \bigl(\,\partial_z+B^\prime_z\,\bigr) \chi_n^{(-)}
+ \frac{1}{\rho} \, \Bigl( (n-1/2) \, \bar \chi_n^{(-)} \chi_n^{(+)}
- (n+1/2)\, \bar\chi_n^{(+)} \chi_n^{(-)} \Bigr)
\Bigr\}\,,
\label{actionnophi}
\end{align}
with $\epsilon$ a positive infinitesimal. 

This action can be interpreted as a three-dimensional field theory
with infinitely many Dirac fermions, labeled by $n \in \bZ\,$. 
These three-dimensional Dirac fields are defined as follows:
\begin{equation}
\eta_n \equiv \vect{ \sqrt{2\pi\rho}\: \chi_n^{(+)} \\ 
 \sqrt{2\pi\rho}\:\chi_n^{(-)}} \,, \quad
\bar{\eta}_n \equiv \Bigl( \sqrt{2\pi\rho}\: \bar\chi_n^{(+)}\,, \, 
\sqrt{2\pi\rho}\: \bar\chi_n^{(-)} \Bigr)\:\gamma^0\,,            
\end{equation}
with $\gamma$-matrices 
\begin{equation}
\gamma^0 \equiv \sigma^2\,,\quad \gamma^1 \equiv \ri \, \sigma^3\,, \quad
\gamma^2 \equiv -\ri \, \sigma^1\,,
\end{equation}
which obey
\begin{eqnarray}
\gamma^0 \gamma^1 \gamma^2 &=& \ri \,, \quad
\{\gamma^\mu, \gamma^\nu\} = 2\,\eta^{\mu\nu}\,,
\end{eqnarray}
for $\mu, \nu =0,1,2$,  and $(\eta^{\mu\nu}) = {\rm diag}\,(1,-1,-1)$.
The action (\ref{actionnophi}) is then given by
\begin{equation}
\Ifermion =  \int_{\widetilde M_3} {\rm d}^3 y \, \sum_{n=-\infty}^{+\infty}
\left\{ \bar \eta_n\: \ri \, \gamma^\mu 
        \bigl(\,\partial_\mu + \widetilde B^\prime_\mu \,\bigr)\, 
        \eta_n + \frac{n}{y^1} \, \bar \eta_n \,\eta_n \right\}\,,
\label{action3d}
\end{equation}
where $t$, $\rho$, $z$ have
been renamed $y^0$, $y^1$, $y^2$, respectively, and
$\widetilde B^\prime_\mu$ has been defined as 
$\widetilde B^\prime_0 \equiv B^\prime_t$, $\widetilde
B^\prime_1 \equiv B^\prime_\rho$, $\widetilde B^\prime_2\equiv B^\prime_z\,$. The relevant three-dimensional
spacetime manifold $\widetilde M_3 = \bR \times \bR_{>0} \times \bR \,$
has no boundary and is topologically equivalent to $\bR^3$.
 
The $n=0$ sector of the theory (\ref{action3d}) describes a massless Dirac
fermion in a background gauge field, whereas the $n\ne 0$ sectors have additional  
position-dependent mass terms. At this moment, there is no need to 
specify the gauge-invariant regularization of the theory, 
one possibility being the use of a spacetime lattice; cf. Refs. \cite{K00, KS02}.

The perturbative quantum field theory based on the action (\ref{action3d})
contains only tree and one-loop diagrams because the
gauge field does not propagate and because there are no fermion 
self-interactions present. 
The effective gauge field action from the 
$n=0$ sector of the field theory (\ref{action3d}) over the spacetime 
$\widetilde M_3$ is directly related to the effective action
from charged massless Dirac fermions over $\bR^3\,$.

The $n=0$ sector of the theory (\ref{action3d}) has therefore the 
same ``parity anomaly'' as the standard $\bR^3$ theory \cite{R84,ADM85,CL89}.
The anomaly manifests itself in a contribution to the effective  
action $\Gamma[\widetilde B^\prime]$ of the form
\begin{equation} 
\int_{\widetilde M_3} \!{\rm d}^3y \,\,s_0\, \pi\, 
\omega_\mathrm{CS}[\widetilde B^\prime_0,
\widetilde B^\prime_1, \widetilde  B^\prime_2]\,,
\label{anom3d}
\end{equation}   
where $\omega_\mathrm{CS}$ is the \CS~density
\begin{equation}
\omega_\mathrm{CS}[B_0, B_1, B_2] \equiv \frac{1}{16\pi^2}\,
\epsilon^{\kappa\lambda\mu}\, \tr \left(B_{\kappa\lambda} B_\mu -
  \tfr{2}{3} B_\kappa B_\lambda B_\mu \right)\,,
\label{omegaCS}
\end{equation}
in terms of the \YM~field strength 
$B_{\kappa\lambda}\equiv
\partial_\kappa B_\lambda -\partial_\lambda B_\kappa + 
[B_\kappa,B_\lambda]$,
with indices running over $0,1,2$, and the \LC~symbol 
$\epsilon^{\kappa\lambda\mu}$ normalized by $\epsilon^{012}=+1\,$. 
The factor $s_0$ in the anomalous term (\ref{anom3d}) is an odd integer 
which depends on the ultraviolet regularization used and  we take 
$s_0=+1\,$. Note that the \CS~integral (\ref{anom3d}) is a topological term, 
i.e., a term which is independent of the metric on $\widetilde M_3\,$. 
The total contribution of the $n\ne 0$ sectors to the effective action
cannot be evaluated easily but is 
expected to lead to no further anomalies; cf. Refs. \cite{K00, KS02}.

The anomalous term (\ref{anom3d}) gives a contribution to the 
four-dimensional effective action $\Gamma[B^\prime]$ of the form
\begin{equation}           
\int_{M} \! {\rm d}^4x \; \frac{\pi}{2\pi\rho}\,
\left\{ \frac{x^1}{\rho} \,
  \omega_\mathrm{CS}[B^\prime_0, B^\prime_1, B^\prime_3] + \frac{x^2}{\rho}\, \omega_\mathrm{CS}[B^\prime_0,
  B^\prime_2, B^\prime_3] \right\}\,,
\label{anom1}
\end{equation}
in terms of the usual Cartesian coordinates $x^\mu$ and the corresponding
four-dimensional gauge fields $B^\prime_\mu(x)$, $\mu=0,1,2,3$, 
and with the definition $\rho^2 \equiv (x^1)^2 + (x^2)^2$. 
In the form written, Eq. (\ref{anom1}) has the same structure as the 
CPT anomaly term (4.1) of Ref.~\cite{K00} for the $\bR^3 \times S^1$ manifold.
The main properties of the anomalous term will be recalled
in the next subsection. 

For completeness, we mention that the CPT anomaly also occurs for 
another type of orientable manifold, $N =\bR \times N_3\,$, 
which is both multiply connected and geodesically complete. 
This particular space manifold $N_3$ has a single wormhole, 
constructed from Euclidean three-space $\bR^3$
by removing the interior of two identical balls   
and properly identifying their surfaces (in this case, without time
shift); cf. Refs.~\cite{Fetal90,V96}.
The space ``defect'' is point-like if the removed balls are
infinitesimally small.
For an appropriate class of gauge fields $B^{\prime\prime}_\mu(x)$, 
the parity anomaly gives directly an action term analogous
to Eq.~(\ref{anom1}), now over the spacetime manifold $N$
and with a dipole structure for the $\omega_\mathrm{CS}$ terms in the integrand. 
The detailed form of this anomalous term is, however, somewhat involved
(see Appendix \ref{sec:single_wormhole})              
and, for the rest of this section, we return to the original manifold
$M = \bR^4\setminus \bR^2$.
\end{widetext}

\subsection{Abelian anomalous term} 
\label{sec:SPabelian}

The gauge field $B_\mu(x)$ will now be restricted to the Abelian Lie subalgebra
$u(1) \subset so(10)$ which corresponds to electromagnetism 
and the resulting real gauge field will be denoted by $A_\mu(x)$. In this
case, the trilinear term of the \CS~density (\ref{omegaCS}) 
vanishes. For the $\phi$-independent gauge fields (\ref{Bprime})
restricted to the $u(1)$ subalgebra,
the anomalous contribution (\ref{anom1}) to the effective 
gauge field action $\Gamma[A^\prime]$ is simply
\begin{equation}
-\frac{1}{8\pi} 
\int_{M} {\rm d}^4x \; \epsilon^{\mu\nu\kappa\lambda}\;
q^\prime_\mu\; \partial_\nu A^\prime_\kappa(x)\: A^\prime_\lambda(x)\,,
\label{anom2}
\end{equation}
with the \LC~symbol $\epsilon^{\mu\nu\kappa\lambda}$
normalized by $\epsilon^{0123}=+1$,
\begin{equation}
q^\prime_\mu \equiv \partial_\mu f^\prime = \frac{g^2}{4\pi}\; 
\left(0\,, \, -\frac{x^2}{\rho^2}\,, \,\frac{x^1}{\rho^2}\,, \, 0 \right)\,,
\label{qprime}
\end{equation}
and $\rho^2 \equiv (x^1)^2 + (x^2)^2$.

Four remarks on the result (\ref{anom2}) are in order. First, 
the action term (\ref{anom2}), with 
$q^\prime_\mu \equiv \partial_\mu f^\prime$,
is invariant under a four-dimensional Abelian gauge 
transformation, $A^\prime_\mu(x) \to A^\prime_\mu(x)+\partial_\mu \,\xi(x)$. 
Second, the Lorentz and time-reversal (T) invariances are broken 
(as is the CPT invariance), because the $q^\prime_\mu$ components in 
the effective action term (\ref{anom2}) are fixed once and for all
to the values (\ref{qprime});  cf.\ Ref.\ \cite{K01wigner}. 
Third, we expect no problems with unitarity and 
causality for the photon field, because the component $q^\prime_0$ vanishes 
exactly; cf.\ Refs.\ \cite{CFJ90,AK01}.
Fourth, the overall sign  of expression (\ref{anom2}) can 
be changed by reversing the direction of the $z$-axis; cf. \ Eq.\ (\ref{anom3d}). 

After a partial integration, the anomalous contribution (\ref{anom2}) can 
be generalized to the following term in the effective action $\Gamma[A]$:
\begin{equation}
\frac{1}{32\pi} \int_{\bR^4} \dx f_{M}(x;A]\; 
\epsilon^{\kappa\lambda\mu\nu} \,
 F_{\kappa\lambda}(x)\, F_{\mu\nu}(x)\,,
\label{anom3}
\end{equation}
with the field strength   
$F_{\mu\nu}(x)\equiv \partial_\mu A_\nu(x) - \partial_\nu A_\mu(x)$
and the integration domain extended to $\bR^4$, which is possible for
smooth gauge fields $A_\mu(x)$. 
(See Refs.\ \cite{CH85,N88,HR01} for a related discussion in the
context of axion electrodynamics.) 
The factor $f_{M}(x;A]$ in the anomalous term (\ref{anom3})
is both a function of the spacetime coordinates $x^\mu$ 
(on which the partial derivative $\partial_\mu$ acts to give $q_\mu$) 
and a gauge-invariant functional of the gauge field $A_\mu(x)$. 
This functional dependence of $f_{M}$ involves, most likely, the gauge field 
holonomies, defined as
$h_{C}[A] \equiv \exp\left(\i \oint_{C} {\rm d}x^\mu \, A_\mu(x)\right)$
for an oriented closed curve $C$; cf. Sec.~4 of Ref.~\cite{K00}.

Note that the functional $f_{M}(x;A]$ in Eq.~(\ref{anom3})
is defined over $\bR^4$ but carries the memory of the 
original (multiply connected) manifold $M$, as indicated by the suffix. 
[The same structure (\ref{anom3}) has also been found for a manifold with 
a single static wormhole; see Appendix \ref{sec:single_wormhole}.] 
Moreover, the absolute value of $f_{M}$ is of the order of the 
fine-structure constant,
\begin{equation}
\left| f_{M}(x;A] \right| = 
\mathrm{O}(\alpha)\,, \quad \alpha \equiv e^2/(4\pi)\,,
\label{absf}
\end{equation}
with the electromagnetic coupling constant $e \propto g\,$. 
The general expression for $f_{M}(x;A]$ is not known, but $f_{M}$ can be
calculated on a case by case basis [that is, the function $f^{\prime}(x)$ for
the gauge field configuration $A^{\prime}$, $f^{\prime\prime}(x)$ for 
$A^{\prime\prime}$, \emph{et cet\-era}].

\section{Models}
\label{sec:Models}

\subsection{Motivation}
\label{sec:Motivation}

The exact calculation of the CPT anomaly is prohibitively difficult 
for two or more punctures or wormholes.  
However, we expect the Abelian anomalies from the individual defects  
(considered in Sec.\ \ref{sec:SingleDefect} and 
Appendix \ref{sec:single_wormhole})
to add up incoherently, at least over large enough scales.
We, therefore, assume that the total anomalous effect of the defects can
be described by a contribution to the effective gauge field action 
$\Gamma[A]$ of the form (\ref{anom3}) but with $f_{M}(x;A]$ 
replaced by a background field $g_1(x)$.  This background field $g_1(x)$
carries the imprint of the topologically nontrivial structure of spacetime
as probed by the chiral fermions. [The original spacetime manifold may, 
of course, have additional structure which does not contribute to the 
CPT anomaly and does not show up in $g_1(x)$.]  

As discussed in the Introduction, we consider in this paper a particular 
type of spacetime foam for which the defects of
three-space are static and have randomly distributed positions and 
orientations. The average distance between the defects can be assumed to be small 
compared to
the relevant scales (set by the photon wavelength, for example)
and the detailed form of $g_1(\vec x)$ is not important for macroscopic
considerations. We, therefore, consider $g_1(\vec x)$ to be a ``random'' 
field and only assume some simple ``statistical'' properties. 
These statistical properties will be specified in 
Sec.\ \ref{sec:RandomBackground}.

It should, however, be clear that the background field $g_1(\vec x)$ is 
not \emph{completely} random. It contains, for example, the small-scale 
structure of the individual anomaly terms. The randomness of $g_1(\vec x)$ 
traces back solely to
the distribution of the static defects whose physical origin is unknown. 
In fact, the aim of the present paper is to establish and constrain   
some general characteristics of these hypothetical defects.

In the rest of this section, we present two concrete models with 
random background fields.
The first model describes the propagation of a single real scalar field
and the second the behavior of the photon field. 
Both models are defined over Minkowski spacetime,
$M=\bR^4$ and $g_{\mu\nu}(x) = \eta_{\mu\nu}\,$.

\subsection{Real scalar field}
\label{sec:ModelScalar}

The scalar model is defined by the action
\begin{align}
I_\mathrm{\, scalar} = &\int_{\bR^4} \dx \, \exp\left[\,g_0(x)\,\right] \no\\ 
&\times\Bigl( \partial_\mu \phi(x)\, \partial^\mu \phi(x) 
       - m^2 \phi(x)^2 \Bigr) \,,
\label{Gammascalar}
\end{align}
where $g_0$ is a real scalar background field of mass dimension
zero. The background field $g_0$ is assumed to be random and further properties
will be discussed in Sec. \ref{sec:RandomBackground}. 

The corresponding  equation of motion reads
\begin{equation} 
(\Box +m^2)\, \phi(x)  = -\partial_\mu g_0(x) \, \partial^\mu \phi(x) \,,
\label{field_eq}
\end{equation}
with the following conventions:
\begin{eqnarray}
\Box &\equiv &\eta^{\mu\nu} \, \partial_\mu \, \partial_\nu\,, \no\\ 
(\eta^{\mu\nu}) &\equiv &{\rm diag} (1,-1,-1,-1)\,.
\label{conventions}
\end{eqnarray}
There is no equation of motion for $g_0$, because $g_0$ is a fixed background 
field [the random coupling constants $g_0(x)$ are quenched variables]. 
The scalar field equation (\ref{field_eq}) will be seen to have the same 
basic structure as the one of the photon model in the next subsection.

For the free real scalar field $\phi^{(0)}$, we define 
\begin{eqnarray}
\phi^{(0)}(x) &= &\frac{1}{(2\pi)^4}\, \int \dint{^4 k} \,
\widetilde \phi^{(0)}(k) \, \e^{-\ri k \cdot x}
\,,\no\\
\widetilde \phi^{(0)}(k) &=&  2\pi\, \delta(k_\mu k^\mu-m^2)\,
\bigl[\, \theta(k^0)  \, \widetilde \phi^{(0)}(\vec k) \no\\[1mm]
&&    + \,\theta(-k^0) \, \widetilde \phi^{(0)}(-\vec k)^* \,\bigr]
\,. \label{freefield}
\end{eqnarray}
The causal Green function $\Delta(x)$ of the \KG~operator, 
\begin{equation}
(\Box + m^2) \, \Delta(x) = \delta^4(x)\,,
\end{equation} 
is given by 
\begin{eqnarray}
\Delta(x)& = &  \frac{1}{(2\pi)^4} \int \dint{^4 k} \, \e^{-\ri k \cdot x} 
    \, \widetilde \Delta(k)\,,\no\\
\widetilde \Delta(k)& = &\frac{-1}{k^2-m^2+\ri\,\epsilon}\,,
\label{scalarprop}
\end{eqnarray}
with the Feynman prescription for the $k^0$ integration contour \cite{IZ80}.

\subsection{Photon field}
\label{sec:ModelPhoton}

The photon model is defined by the action
\begin{align}
I_\mathrm{\, photon} = &-\frac{1}{4} \int_{\bR^4} \dx 
  \Bigl( F_{\mu\nu}(x)  F^{\mu\nu}(x) \no\\
  &+ \, g_1(x)\,  F_{\kappa\lambda}(x)  \widetilde F^{\kappa\lambda}(x)
  \Bigr) \,, \label{Gammaphoton}
\end{align}
where the Maxwell field strength tensor $F_{\mu\nu}$ and its dual 
$\widetilde F^{\kappa\lambda}$ are given by
\begin{eqnarray}
F_{\mu\nu} &\equiv& \partial_\mu A_\nu -\partial_\nu A_\mu\,,\no\\
\widetilde F^{\kappa\lambda} &\equiv& 
\tfr{1}{2}\;\epsilon^{\kappa\lambda\mu\nu}\, F_{\mu\nu} \,,
\end{eqnarray}
with $\epsilon^{\kappa\lambda\mu\nu}$ the \LC~symbol 
normalized by $\epsilon^{0123}=+1\,$. 

The random (time-independent) background field $g_1$ in the action 
(\ref{Gammaphoton}) is supposed to mimic the anomalous 
effects of a multiply connected (static) spacetime foam, 
generalizing the result (\ref{anom3}) for a single puncture or wormhole.
Following Eq.~(\ref{absf}), the amplitude of the random background field $g_1$ 
is assumed to be of order $\alpha$. The typical length scale over which $g_1$ 
varies will be denoted by $l_{\rm foam}$     
and further properties will be discussed in Sec.\ \ref{sec:RandomBackground}. 
Note that models of the type (\ref{Gammaphoton}) 
have been considered before, but, to our knowledge, only for 
coupling constants varying smoothly over cosmological 
scales; cf. Refs.~\cite{CFJ90,KLP02}.

The equations of motion corresponding to the action (\ref{Gammaphoton}) 
are given by
\begin{equation}
\Box \, A^\nu(x) = - \partial_\mu g_1(x)\, \widetilde F^{\mu\nu}(x)\,,
\label{photon_field_eq_pos}
\end{equation}
provided the Lorentz gauge is used,
\begin{equation}
\partial^\mu A_\mu(x)=0\,.
\label{LorentzGauge}
\end{equation}
The random coupling constants $g_1(x)$ in the action (\ref{Gammaphoton}) 
are considered to be quenched variables 
and there is no equation of motion for $g_1$.
As mentioned above, the basic structure of Eq.~(\ref{photon_field_eq_pos}) 
equals that of Eq.~(\ref{field_eq}) for the scalar model. 

The general solution of the free field equation, $\Box A_\mu =0$, is
\begin{eqnarray}
A^{(0)}_\mu(x) &= &\frac{1}{(2\pi)^4} \int \dint{^4k} \widetilde
A^{(0)}_\mu(k)\, \e^{-\ri k \cdot x} \,, \no\\[1mm]
\widetilde A^{(0)}_\mu(k) &=&  2\pi\, \delta(k_\mu k^\mu)\,
\bigl[\,  \theta(k^0)  \, \widetilde A^{(0)}_\mu(\vec k) \no\\
&&         +\, \theta(-k^0) \,\widetilde A^{(0)}_\mu(-\vec k)^*  \,\bigr]\,,    
\end{eqnarray}
with
\begin{equation}
k^\mu   \widetilde A^{(0)}_\mu(k) =0\,,
\label{LorentzGauge2}
\end{equation}
due to the gauge condition (\ref{LorentzGauge}).
The Feynman propagator \cite{IZ80} for the free field
is  simply $\Delta_{\mu\nu}(x) \equiv \i \, \eta_{\mu\nu}\, \Delta(x)$, 
with $\Delta(x)$ given by Eq.~(\ref{scalarprop}) for $m=0\,$.

\section{Random background fields}
\label{sec:RandomBackground}

The two models of the previous section have random background fields $g_0$
and $g_1$. For simplicity, we assume the same basic properties for 
these two random background fields and denote $g_0$ and $g_1$
collectively by $g$ in this section.

\subsection{General properties}
\label{sec:RandomGeneral}

The assumed properties of the background field $g(x)$ are the following:
\begin{enumerate}
\item $g$ is time-independent, $g=g(\vec x)$,
\item $g$ is weak, $|g(\vec x)|\ll 1$,
\item the average of $g(\vec x)$ vanishes in the large volume limit, 
\item $g(\vec x)$ varies over length scales which are     
      small compared to the considered wavelengths of the scalar field $\phi$ 
      and photon field $A_\mu \,$,
\item the autocorrelation function of $g(\vec x)$ is finite and isotropic,
      and drops off ``fast enough'' at large separations 
      (see Secs.~\ref{sec:RandomPhase} and   \ref{sec:Dispersion} below). 
\end{enumerate}
\par
The random background field $g_1(\vec x)$ for the photon case is considered to 
incorporate the effects of a multiply connected, static spacetime foam
(cf. Sec.\ \ref{sec:Motivation}) 
and assumption 5 about the lack of long-range correlations
can perhaps be relaxed. Indeed, long-range correlations could arise from 
permanent or transient wormholes in spacetime; cf. Refs.\ \cite{GH93,V96}. 
But, for simplicity, we keep the five assumptions as listed above.

\subsection{Example}
\label{sec:BackgroundExample}

A specific class of random background fields $g(\vec x)$ can 
be generated by superimposing copies of a localized, square-integrable 
shape $h(\vec x)$  with random displacements.
The displacements $\vec x_n$ are uniformly distributed over a ball of
radius $R$ embedded in 
$\bR^3$ and have average separation $a$. The number of elementary
shapes is then given by $N=(4\pi/3) R^3/a^3$.  

Concretely, we take
\begin{equation}
g_N(\vec x) =  \alpha \, \sum_{n=1}^N \epsilon_n\, h(\vec x-\vec x_n)\,,
\label{gN}
\end{equation}
where the numbers $\epsilon_n=\pm 1$ are chosen randomly, so that
$g_N(\vec x)$ vanishes on average. The background field $g(\vec x)$ is 
defined as the infinite volume limit,
\begin{equation}
 g(\vec x) = \lim_{N\to \infty} \, g_N(\vec x)\,,
\end{equation}
where $a$ is held constant.

The mean value and autocorrelation function of $g$ are defined by
\begin{eqnarray} 
\langle g \rangle \!&\equiv&\! \lim_{R\to \infty} \frac{1}{(4\pi/3) R^3}
\!\int\limits_{|\vec x|<R} \! \dvx g(\vec x) \,, \\[1mm]
C(\vec x) \!&\equiv&\! \lim_{R\to \infty} \frac{1}{(4\pi/3) R^3}
\!\int\limits_{|\vec y|<R} \!\dvy  g(\vec y) \, g(\vec y + \vec x)\, . 
\label{correl}
\end{eqnarray}
In the limit $R\to \infty$, we have perfect statistics ($N\to\infty$) and find
\begin{eqnarray}
\langle g \rangle  &=& 0\,,\\
C(\vec x) &=& \alpha^2\:a^{-3}\, \int_{\bR^3} \dvy \,
              h(\vec y) \, h(\vec y + \vec x)\,. 
\label{Cexample}
\end{eqnarray}
\par
\begin{figure*}[p]         
\includegraphics[width=14cm]{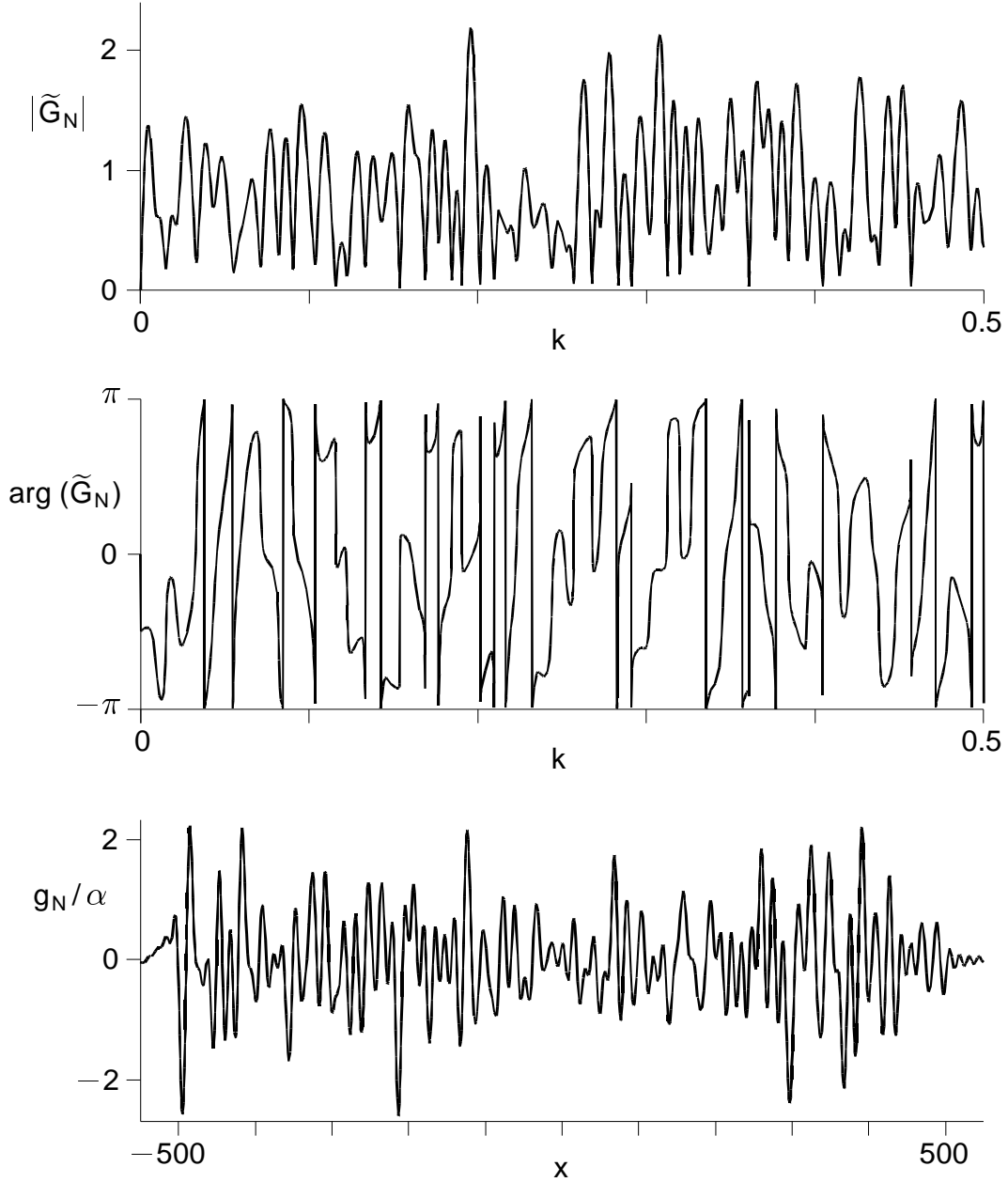} 
\vspace*{2\baselineskip} 
\caption{The top and middle panels show the momentum-space function
$\widetilde G_N(k)$, defined by the one-dimensional version of Eq.~(\ref{tildeGN}), 
for $N=100$ and $R=500$.
The bottom panel shows the corresponding position-space function $g_N(x)$,  
defined by the one-dimensional version of Eq.~(\ref{gN}), 
for the profile function     
$h(x) = (20/9)\left(\,\sin\, x/2 - \sin\, x/20 \,\right)/x$.  
The Fourier transform  $\widetilde{h}(k)$ of this particular profile function 
is nonzero and constant for
$|k| \in [k_\mathrm{low}, k_\mathrm{high}]$, with 
$ k_\mathrm{high} = 10 \:k_\mathrm{low} >0$, and zero otherwise.} 
\label{fig:GR}
\end{figure*}

The Fourier transforms of $h$ and $g_N$ are given by
\begin{subequations} 
\begin{eqnarray}
\widetilde h(\vec k) &=&\int_{\bR^3} 
\dvx \e^{-\ri
  \vec k \cdot \vec x}\, h(\vec x)\,,\\[2mm]
\widetilde g_N(\vec k) &=& \alpha \, \sum_{n=1}^N \epsilon_n\,  \int_{\bR^3}  
 \dvx \e^{-\ri \vec k \cdot \vec x}\, h(\vec x-\vec x_n) \no\\[2mm]
&=& \alpha \, \sqrt{N}\;\, \widetilde h(\vec k)\: \widetilde G_N(\vec k) \,,
\end{eqnarray}
\end{subequations}
with
\begin{equation}
\widetilde G_N(\vec k) \equiv 
\frac{1}{\sqrt{N}}\;\sum_{n=1}^N \epsilon_n\,  \e^{-\ri \vec k \cdot \vec x_n} \,.
\label{tildeGN}
\end{equation}
The function (\ref{tildeGN}) can  be decomposed into an absolute value and a 
phase factor. The absolute value
fluctuates around $1$, as follows from the expression 
\begin{equation}
\left|\widetilde G_N (\vec k)\right|^2  = 
 1 + \frac{1}{N}\, \sum_{n\ne m} \epsilon_n\epsilon_m \,
 \e^{-\ri \vec k \cdot (\vec x_n-\vec x_m)}\,,
\label{GNabs}
\end{equation}
where the double sum on the \rhs~scatters around $0$. 
The fluctuation scale of $|\widetilde G_N (\vec k)|$ is of order $2\pi/R$ and 
the same holds for the phase of $\widetilde G_N (\vec k)$; see Fig.\ \ref{fig:GR}.

\subsection{Random phase assumption}
\label{sec:RandomPhase}
For a given random background field $g(\vec x)$ over $\bR^3$, 
we define the truncated Fourier transform $\widetilde g_R(\vec k)$ by
\begin{equation}
\widetilde g_R(\vec k) \equiv \int\limits_{|\vec x|<R} \dvx 
                          \e^{-\ri\vec k\cdot \vec x}\, g(\vec x)\,,
\end{equation}
with an implicit dependence on the origin of the sphere chosen.
This function $\widetilde g_R$ can be parametrized as follows: 
\begin{equation}
\widetilde g_R(\vec k) =\sqrt{(4\pi/3) R^3}\,\, \widetilde H(\vec k)\, \widetilde
  G_R(\vec k)\,,
\label{tgRparam}
\end{equation}
where the real function $\widetilde H(\vec k)$  is obtained from
the finite autocorrelation function (\ref{correl}), 
\begin{equation}
C(\vec x) = \frac{1}{(2\pi)^3} \int \dint{^3\vec k} \, \e^{\ri
  \vec k \cdot \vec x} \, {\widetilde H(\vec k)^2}\,,
\label{F2_1}
\end{equation}
with the sign of $\widetilde H(\vec k)$ chosen to make the function as 
smooth as possible (i.e., without cusps).   
Because the background field $g(\vec x)$ is real, we have also
$\widetilde H(-\vec k)\:\widetilde G_R(-\vec k)= 
\widetilde H(\vec k)\:\widetilde G_R(\vec k)^*\,$.  

Motivated by the results of the preceding subsection, we assume 
$\widetilde G_R(\vec k)$ to vary over momentum intervals of the order of 
$2\pi/R$ and $|\widetilde G_R|$
to have unit mean value.  Furthermore, $\widetilde H(\vec k)^2 $ 
is taken to be smooth on scales of the order of $2\pi/R$, 
provided $R$ is large enough.  
For the isotropic case $\widetilde H(\vec k)=\widetilde H(k)$, 
with $k\equiv |\vec k|$, and large enough $R$,
we explicitly assume the inequality 
\begin{equation}
\left|\, \int_0^\infty \dint{k} \cos (k R)\, \widetilde H(k)^2 \,\right| 
\ll \int_0^\infty
\dint{k} \widetilde H(k)^2\,,
\label{Hsmooth}
\end{equation}
with a finite positive number on the \rhs.

Now suppose that we have to evaluate a double integral of the form
\begin{equation}
\int \dint{^3\vec k} \! \int \dint{^3\vec p} f(\vec p, \vec k)\,
\widetilde G_R(\vec k) \, \widetilde G_R(\vec p)\,,
\end{equation}
where $f$ is a function which is approximately constant over 
momentum intervals of the order of
$2\pi/R\,$. Due to the rapid phase oscillations of $\widetilde G_R$, a significant
contribution can arise only for $\vec k \approx -\vec p\,$ and
$\widetilde G_R(\vec k) \, \widetilde G_R(\vec p)$ 
can effectively be replaced by a smeared delta-function. 

The specific form of this smeared delta-function 
does not matter in the limit $R\to \infty$ and we simply choose 
\begin{equation}
\delta_R^3(\vec k) \equiv \prod_{j=1}^3 \,\frac{\sin(k_j R)}{\pi k_j}\;.
\label{deltaRdef}
\end{equation}
We then have the result
\begin{align}
&\int \dint{^3\vec k} \! \int \dint{^3\vec p} f(\vec p, \vec k)\,
\widetilde G_R(\vec k) \, \widetilde G_R(\vec p) \no\\
&\approx 
{\cal N}\,\int \dint{^3\vec k} \! \int \dint{^3\vec p} f(\vec p, \vec k)\,
\delta_R^3(\vec k + \vec p)\,,
\label{GGrep}
\end{align}
with a normalization factor $\cal N$ to be determined shortly.
On the \rhs~of Eq.~(\ref{GGrep}), 
the absolute values of $\widetilde G_R$ have been replaced by their average value 
$1$, since $f(\vec p, \vec k)$ is assumed to be slowly varying. 

In order to determine the normalization factor $\cal N$, reconsider
the autocorrelation function (\ref{correl}). Using Eqs.\ (\ref{tgRparam}) and 
(\ref{deltaRdef}), the autocorrelation function can be written as follows: 
\begin{align}
C(\vec x) = &\lim_{R\to \infty} \frac{1}{
  (2\pi)^3} \int \dint{^3\vec k} \dint{^3\vec p}\, \delta^3_R
  (\vec k+\vec p) \,\, \e^{\ri \vec p\cdot \vec x}\no\\ 
&\times \widetilde H(\vec k)\, \widetilde
  H(\vec p) \,\, \widetilde G_R(\vec k)\,\widetilde G_R(\vec p)\,. 
\label{F2_2}
\end{align}
Since there is already a delta-function present in the integrand
of (\ref{F2_2}), the product 
$\widetilde G_R(\vec k)\,\widetilde G_R(\vec p)$ can simply be replaced by 
${\cal N}\, \delta_R^3 (\vec k+\vec p)$, which gives the correct behavior for 
$\vec k \approx -\vec p\,$. With
\begin{equation}
 \delta^3_R(\vec k+\vec p) \, \delta^3_R(\vec k+\vec p)
\to \frac{R^3}{\pi^3} \,\,\delta^3(\vec k+\vec p)\,, 
\end{equation}
for $R\to \infty$, Eqs.~(\ref{F2_1}) and (\ref{F2_2}) are consistent if
\begin{equation}
{\cal N}=(\pi/R)^3 \,.
\label{calN}
\end{equation}
This fixes the normalization factor $\cal N$ of 
Eq.~(\ref{GGrep}), which will be used extensively in the next section.

\begin{widetext}
\section{Dispersion laws}
\label{sec:Dispersion}

\subsection{Scalar model}
\label{sec:ScalarDispersion}

We first turn to the scalar model (\ref{Gammascalar}), as a preliminary for 
the calculation of the photon model in the next subsection. In order to have
a well defined Fourier transform of $g_0$, the system is put inside a sphere
of radius $R$ with free boundary conditions for the real scalar field $\phi$.
In momentum space, the scalar field equation (\ref{field_eq}) becomes
\begin{equation}
(k^2-m^2) \, \widetilde \phi(k) = - \frac{1}{(2\pi)^4} \int \dint{^4 q}
 \widetilde g_{0R}(q)\, q_\mu\, (k^\mu-q^\mu)\,  \widetilde \phi(k-q)\,.
\label{field_eq2}
\end{equation}
According to Eq.~(\ref{tgRparam}), the momentum-space function 
$\widetilde g_{0R}$ has the form 
\begin{equation}
  \widetilde g_{0R}(k)\equiv 2\pi \delta(k^0) \,
  \sqrt{ (4\pi/3)\, R^3}\,\,  \widetilde H_0(|\vec k|)\, \widetilde G_{0R}(\vec
  k)\,,
\label{tgRparam2}
\end{equation}
where $\widetilde G_{0R}$ is assumed to have the statistics
properties discussed in Sec.~\ref{sec:RandomPhase}. Furthermore, we have
assumed isotropy of $\widetilde H_0$; see Sec.\ \ref{sec:RandomGeneral}.

We will now show that the main effect of the random background field 
$g_0 (\vec x)$ can be 
expressed in the form of a modified dispersion law for the scalar modes, 
at least for large enough wavelengths. 
The basic idea is to expand the solution of
Eq.~(\ref{field_eq2}) perturbatively to second order in $g_0$ (there is no
contribution at first order) and to compare
it with the first-order solution of a modified field equation,
\begin{equation}
(k^2-m^2) \, \widetilde \phi(k) = A \bigl(|\vec k| \bigr)\; \widetilde \phi(k)\,.
\label{A_field_eq}
\end{equation}
The dispersion law will then be given by
\begin{equation} \label{generaldisplaw}
\omega^2= m^2+|\vec k|^2 + A \bigl( |\vec k| \bigr) \,,
\end{equation}
with $A$ expressed in terms of the random background field $g_0 (\vec x)$.

In the limit $R\to \infty$, the perturbative second-order contribution
to $\widetilde \phi(k)$ reads
\begin{align}
\widetilde \phi^{(2)}(k) = &\lim_{R\to\infty}\, \frac{1}{(2\pi)^8}\:\widetilde\Delta(k) 
  \int \dint{^4 q}  \widetilde g_{0R}(q)\, \, q_\mu \, (k-q)^\mu \,
  \widetilde \Delta(k-q)  \nonumber \\[1ex]
& \times \int \dint{^4 p} \widetilde g_{0R}(k-p-q)\,\, (k-p-q)_\nu
  \: p^\nu \, \widetilde \phi^{(0)}(p)\nonumber\\[1ex]
=& - \lim_{R\to\infty}  \, \widetilde \Delta(k)\,  \frac{(4\pi/3) R^3}{(2\pi)^6}\, 
  \int \dint{^3 \vec q} \int
  \dint{^3 \vec p} \, \widetilde H_0(|\vec q|)\,\, \widetilde
  H_0(|\vec k-\vec p-\vec q|)\,\, \widetilde G_{0R}(\vec q)  \, \nonumber \\[1ex]
& \times  \, \widetilde G_{0R}(\vec
  k-\vec p-\vec q)\,\,  \widetilde \Delta(k^0, \vec k-\vec q)
    \left( \vec q \cdot(\vec q-\vec 
  k)\right) \left( (\vec k-\vec p-\vec q)\cdot \vec p \, \right)\, 
  \widetilde\phi^{(0)}(k^0, \vec p) \,,
\label{tildephi2}
\end{align} 
where Eq.~(\ref{tgRparam2}) has been used.
According to Eqs. (\ref{GGrep}) and  (\ref{calN}), we can replace 
the product $\widetilde G_{0R}\, \widetilde G_{0R}$ by $(\pi/R)^3 \, \delta^3_R$ 
and take the limit $R\to \infty$,
\begin{align}
\widetilde \phi^{(2)}(k) &=  \frac{1}{12\,(2\pi)^2}  \; \widetilde\Delta(k)\:
  \int \dint{^3 \vec q} \, \widetilde H_0(|\vec q|)^2 \, 
\widetilde \Delta(k^0, \vec k-\vec q) \, \left( \vec q \cdot(\vec q-\vec
  k)\right) \left( \vec q\cdot \vec k \,\right)\, \widetilde \phi^{(0)}(k) \,.
\label{phi2k}
\end{align}
\par
At the given perturbative order,   
$k_\mu k^\mu-m^2$ can be put to zero in the integrand of Eq.~(\ref{phi2k}),   
because of the delta-function contained in the 
free field $\widetilde \phi_0(k)$ of Eq.~(\ref{freefield}).
By comparison with the first order solution of Eq.~(\ref{A_field_eq}), 
the operator $A$ is identified as
\begin{equation}
A \bigl( |\vec k| \bigr) =  
\frac{1}{12\,(2\pi)^2} \,  \int \dint{^3 \vec q} \, \widetilde
H_0(|\vec q|)^2 \; \frac{ \bigl( \vec q \cdot(\vec q-\vec k)\bigr) 
  \bigl( \vec q\cdot \vec k \,\bigr)}
  {2\vec k \cdot \vec q - |\vec q|^2 +\ri\,\epsilon} \,.
\label{A_scalarintegral} 
\end{equation}
\par
Following the discussion of Sec.\ \ref{sec:RandomGeneral}, we now assume that  
$\widetilde H_0(|\vec q|)$ vanishes for momenta $|\vec q|<q_{\rm low}$ 
(cf. Fig.~\ref{fig:GR}) and that $|\vec k| < q_{\rm low}/2$.
Performing the angular integrals in Eq.~(\ref{A_scalarintegral}), one finds
\begin{equation}
A \bigl( |\vec k| \bigr) = \frac{1}{24\pi} \, \int_0^\infty \dint{q} q^2\,
\widetilde H_0(q)^2 \left( \frac{1}{2}\,q^2 + \frac{q^3}{8|\vec k|} \:
 \ln \left|  \frac{q-2|\vec k|}{q+2|\vec k|} \right|\, \right) \,,
\label{A_scalar}
\end{equation}
where the lower limit of the integral is effectively $q_{\rm low}$. 
Because of the large-wavelength assumption $|\vec k| < q_{\rm low}/2$,
the same result is obtained if the causal Green 
function $\widetilde\Delta$ in the first integral of Eq.~(\ref{tildephi2}) is 
replaced by, for example, the retarded or the advanced Green function.

Next, the logarithm of Eq.~(\ref{A_scalar}) is expanded in terms of 
$|\vec k|/q$, which gives 
\begin{equation}
A \bigl( |\vec k| \bigr)  =  
\frac{1}{24 \pi} \, \int_{0}^{\infty} \dint{q} \, \widetilde H_0(q)^2 
\left( -\frac{2}{3} \,q^2\, |\vec k|^2 -
  \frac{8}{5} \, |\vec k|^4 \right) 
  + \mathrm{O}\left(|\vec k|^6/q_{\rm low}^4\right)\,.
\label{A_scalar_exp}
\end{equation}
This result can also be written as
\begin{equation}
A \bigl( |\vec k| \bigr)  = 
-\frac{\pi}{18} \, C_0(0) \, |\vec k|^2 - \frac{2\pi}{15} 
\left(\, \int_0^\infty \dint{x}  x \, C_0(x)\right)
 |\vec k|^4+ \mathrm{O}\left(|\vec k|^6/q_{\rm low}^4\right) \,,
\label{A}
\end{equation}
in terms of the autocorrelation function (\ref{F2_1}),
which is isotropic because $\widetilde H_0$ is, 
\begin{equation}
C_0 = C_0\bigl(|\vec x|\bigr) = 
      \frac{2}{(2\pi)^2} \, \int_0^\infty \dint{q} q^2\;
      \frac{\sin \,q |\vec x|}{q |\vec x|}  \; \widetilde H_0(q)^2  \,.  
\label{C0}
\end{equation}
In order to make the identification of the $|\vec k|^4$ prefactor in
Eq.~(\ref{A}), it has been assumed that the behavior of $\widetilde H_0(q)^2$ is
sufficiently smooth; cf. Eq.\ (\ref{Hsmooth}).

From Eqs.~(\ref{generaldisplaw}) and (\ref{A}), the dispersion law of the
scalar is  
\begin{equation}
\omega^2 = m^2\, +(1-a_0) \, |\vec k|^2  \, -b_0\, |\vec k|^4\, +
\,\dots\,,
\label{scalar_disp_law}
\end{equation}
where the dots stand for higher order contributions and the positive coefficients
$a_0$ and $b_0$ are given by
\begin{equation}
a_0 \equiv \frac{\pi}{18} \, C_0(0) \,, \quad
b_0 \equiv \frac{2\pi}{15} \int_0^\infty \dint{x}  x\, C_0(x)\,.
\end{equation}
The isotropic autocorrelation function $C_0(x)$ is defined by 
Eq.~(\ref{correl}) with $g$ replaced by the random background field
$g_0$ from the action (\ref{Gammascalar}).

The main purpose of the scalar calculation is to prepare the way for
the photon calculation in the next subsection, but let us comment
briefly on the result found.
Setting the scalar mass $m$ to zero in the action (\ref{Gammascalar}), 
our calculation gives no additional
mass term in the dispersion law (\ref{scalar_disp_law}). 
It is, however, not clear what the random background field $g_0$ of the
model (\ref{Gammascalar}) really has to do with a (static) spacetime 
foam. The propagation of an initially massless scalar could very well
be strongly modified in a genuine spacetime foam; cf.\ Ref.~\cite{HPP80}.

\subsection{Photon model}
\label{sec:PhotonDispersion} 

To obtain the dispersion law for the photon model (\ref{Gammaphoton}),
we proceed along the same
lines as for the scalar case. Again, the system is put inside a sphere
of radius $R$, so that the truncated Fourier transform $\widetilde g_{1R}$
occurs in the momentum-space field equation,
\begin{equation}
k^2 \widetilde A^\nu(k) = - \frac{1}{(2\pi)^4} \, \int \dint{^4q}
\widetilde g_{1R}(q) \, \epsilon^{\mu\nu\kappa\lambda}\, q_\mu\, (k-q)_\kappa
\, \widetilde A_\lambda(k-q)\,.
\label{photon_field_eq}
\end{equation}
For this truncated  background field $\widetilde g_{1R}$, we assume a form
analogous to Eq.~(\ref{tgRparam2}),
\begin{equation}
  \widetilde g_{1R}(k)\equiv 2\pi \delta(k^0) \,
  \sqrt{(4\pi/3)\, R^3}\,\,  \widetilde H_1(|\vec k|)\, \widetilde G_{1R}(\vec
  k)\,,
\label{tgRparam3}
\end{equation}
where $\widetilde G_{1R}$ is a random function of the type
discussed in Sec.~\ref{sec:RandomPhase}.

The second-order contribution to the perturbative solution of
Eq.~(\ref{photon_field_eq}) is given by
\begin{align}
\widetilde A^{(2)\nu}(k) &=  \lim_{R\to \infty} \,
\frac{1}{(2\pi)^8} \, \widetilde \Delta(k) \,
\int \dint{^4q} \widetilde g_{1R}(q)\, \epsilon^{\mu\nu\kappa\lambda}\,
q_\mu\, (k- q)_\kappa\,\,
\widetilde \Delta(k-q)\nonumber \\
& \quad \times \,\int \dint{^4p}\, \widetilde g_{1R}(k-p-q)\,
\epsilon_{\alpha \beta \gamma \lambda} \, (k-p-q)^\alpha\, p^\beta\,
\widetilde A^{(0)\gamma} (p)\,.
\end{align}
Using Eq.~(\ref{tgRparam3}) and replacing the product 
$\widetilde G_{1R}\, \widetilde G_{1R}$ by $(\pi/R)^3\, \delta^3_R\,$, 
one finds in the limit $R\to \infty$:
\begin{equation}
\widetilde A^{(2)\nu}(k) = - \widetilde \Delta(k) \, B^{\nu}{}_\gamma(k)\,
                       \widetilde A^{(0)\gamma}(k)\,,
\label{A2nu}
\end{equation}
with
\begin{equation}
B^{\nu}{}_\gamma (k) \equiv \frac{1}{12\,(2\pi)^2}\, \int \dint{^3\vec q}
\widetilde H_1 (|\vec q|)^2\, \frac{1}{(k-q)^2 +\ri\,\epsilon}\;\,
\delta^\mu_{[\alpha} \delta^\nu_\beta \delta^\kappa_{\gamma]}\, \, 
q_\mu \,(k-q)_\kappa \,\,q^\alpha \,k^\beta\,,
\end{equation}
where $q^0 \equiv 0$ and the square brackets around the indices 
$\alpha\beta\gamma$ denote antisymmetrization with unit weight.

Since $\widetilde A^{(0)\gamma}(k)$ in Eq.~(\ref{A2nu})
contains a factor $\delta(k_\mu k^\mu)$
and furthermore obeys the Lorentz gauge condition
(\ref{LorentzGauge2}), we have 
\begin{equation}
B^0{}_\nu \,\widetilde A^{(0)\nu}=0\,,\quad
B^{n}{}_\mu \, \widetilde A^{(0)\mu} = \hat B^n{}_m \, \widetilde A^{(0)m} \,,
\end{equation}
with                
\begin{align}
\hat B^n{}_m \bigl( \vec k \bigr)  &= \frac{1}{12\,(2\pi)^2} 
    \left( \delta_m^n + \frac{k^n k_m}{|\vec k|^2}\right)
    \int \dint{^3\vec q} \widetilde H_1(|\vec q|)^2\;
    \frac{\bigl(\vec q \cdot \vec k \bigr)^2}
         {2\vec k \cdot \vec q - |\vec q|^2 +\ri\,\epsilon} 
=  \left( \delta_m^n + \frac{k^n k_m}{|\vec k|^2}\right) 
          \hat B \bigl( |\vec k| \bigr) \,,
\label{hatBnm}
\\
\intertext{and}
\hat B \bigl(|\vec k| \bigr) &\equiv \frac{1}{24\pi} \, \int_0^\infty \dint{q} \,
    q^2 \widetilde H_1(q)^2\, \left( \frac{1}{2}\,q^2 + \frac{q^3}{8|\vec k|}\:
    \ln \left| \frac{q-2|\vec k|}{q+2|\vec k|}\right|\, \right)\,, \label{hatB}
\end{align}
where the indices $m$ and $n$ run over $1,2,3$.
In order to perform the angular integrals in Eq.~(\ref{hatBnm}), we have
again assumed that $\widetilde H_1(|\vec q|)$ 
vanishes for momenta $|\vec q|<q_{\rm low}$ (cf. Fig.~\ref{fig:GR})
and that $|\vec k| < q_{\rm low}/2\,$. Making an expansion in    
$|\vec k|/q$ for the logarithm in Eq.~(\ref{hatB}), we obtain
\begin{equation}
\hat B \bigl( |\vec k| \bigr) = \frac{1}{24\pi} \, \int_{0}^{\infty} \dint{q} \,
\widetilde H_1(q)^2\,  \left( -\frac{2}{3}\, q^2 \, |\vec k|^2\, -
  \frac{8}{5}\, |\vec k|^4 +\, \dots \right)\,,
\label{hatBexp}
\end{equation}
which is equivalent to the scalar result (\ref{A_scalar_exp}). 
\end{widetext}

The modified field equations in momentum space are thus
\begin{eqnarray}
k^2 \, \widetilde A^0(k)=0\,,\quad
k^2 \widetilde A^m(k)   = \hat B^m{}_n \bigl( \vec k \bigr) \,\widetilde A^n(k)\,.
\end{eqnarray}
It follows that the dispersion law for the scalar and longitudinal modes 
remains unchanged,
\begin{eqnarray}
\omega^2 &=&  |\vec k|^2\,, 
\label{omega2longitudinal}
\end{eqnarray}
whereas the one for the transverse modes is modified,
\begin{eqnarray}
\omega^2 &=&  |\vec k|^2 + \hat B \bigl(|\vec k| \bigr) \,.
\label{omega2transverse} 
\end{eqnarray}
Of course, only the transverse modes are physical;
see, e.g., Sec. 3.2 of Ref. \cite{IZ80}. 
From Eqs.\ (\ref{hatBexp}) and (\ref{omega2transverse}), 
the dispersion law for the (transverse) photons is 
\begin{equation}
\omega^2 = (1-a_1) \, c^2\, |\vec k|^2  \, -b_1\, c^2\, |\vec k|^4\, + \,\dots\,,
\label{photon_disp_law}
\end{equation}
where the bare light velocity $c$ has been restored and the positive coefficients
$a_1$ and $b_1$ are given by
\begin{equation}
a_1 \equiv \frac{\pi}{18} \, C_1(0) \,, \quad
b_1 \equiv \frac{2\pi}{15} \int_0^\infty \dint{x}  x\, C_1(x)\,.
\label{a1b1}
\end{equation}
The isotropic autocorrelation function $C_1(x)$ is defined by 
Eq.~(\ref{correl}) with $g$ replaced by the random coupling constant
$g_1$ from the action (\ref{Gammaphoton}).

According to the discussion in Secs.\ \ref{sec:SPabelian} and 
\ref{sec:ModelPhoton}, the random background field $g_1(\vec x)$
has a typical amplitude of the order of the fine-structure constant
$\alpha$ and a typical length scale of the order of $l_{\rm foam}$. 
The coefficients (\ref{a1b1}) may therefore be written as 
\begin{equation}
a_1 \equiv \alpha^2 \, \gamma_1\,, \quad b_1 \equiv \alpha^2 \: 
l_{\rm foam}^2\,,
\label{lfoamdef}
\end{equation}
with a positive constant $\gamma_1$. 
This last equation defines, in fact, the length scale $l_{\rm foam}$.  
For the example background  
field of Sec.\ \ref{sec:BackgroundExample} with isotropic profile function 
$h = h(|\vec x|)$, one has 
$l_{\rm foam} \propto l_h\, (l_h / a)^{3/2}$,  where $l_h$ is the typical
length scale of the profile function $h$; cf. Eq.~(\ref{Cexample}).
For this type of background field, the ratio $l_{\rm foam} /a$ can be
large if $l_h \gg  a \,$. In Appendix \ref{sec:wormhole_gas},                  
another example background field is discussed,     
which is especially tailored to the case of permanent wormholes 
(average distance $a$ between the different wormholes, 
effective transverse width $2\, l_h$ for the individual wormholes,
and long distance $d$ between the individual wormhole mouths). 
The result for $l_{\rm foam}$ is again found to be proportional to 
$l_h\, (l_h / a)^{3/2}$ and only weakly dependent on the individual
scale $d$.   

The dispersion law  (\ref{photon_disp_law}) 
violates Lorentz invariance but not CPT invariance. 
There is no birefringence, as the background field $g_1(\vec x)$ is assumed 
to have no preferred direction. Note also that motion of the detector relative 
to the preferred frame defined implicitly by the static background $g_1(\vec x)$
would bring in some anisotropy but, at the order considered, no birefringence
(the two modes propagate identically in the instantaneous rest frame
of the detector).

The modifications of the photon dispersion law
found in Eq.~(\ref{photon_disp_law}) are rather mild
(see Ref.\ \cite{L03} for some general remarks on the absence of odd
powers of $|\vec k|$ in modified dispersion laws).
These modifications are not unlike those of Ref.~\cite{HPP80} for a
simply connected,  
nonstatic spacetime made-up of many $S^2 \times S^2$ 
``gravitational bubbles.''  
More drastic changes in the photon dispersion law have, for example, 
been found in certain loop quantum gravity calculations \cite{GP99,AMU02}.
Compared to these calculations (which have the Planck length as the 
fundamental scale),  ours is relatively straightforward,
the only prerequisite being a multiply connected topology 
which is then probed by the chiral fermions of the \SM; 
see Sec.\ \ref{sec:conclusion} for further discussion.

\section{Experimental limit} 
\label{sec:explimit}

In the previous section, we have derived the photon dispersion law for
the simple model (\ref{Gammaphoton}).  Using the definitions (\ref{lfoamdef}) 
and considering $\alpha$ to be parametrically small, we have 
the following expression for the
quadratic and quartic terms in the dispersion law (\ref{photon_disp_law}):
\begin{equation}
\omega^2 \approx c_\mathrm{ren}^2 \, k^2 
 - c_\mathrm{ren}^2 \, \alpha^2 \, l_{\rm foam}^2 \,k^4\,, 
\label{disp_foam}  
\end{equation}
with the renormalized light velocity 
\begin{equation}
 c_\mathrm{ren} \equiv  \,c\,\sqrt{1-\alpha^2 \, \gamma_1} 
\label{cren}
\end{equation}
and the simplified notation $k\equiv |\vec k|\,$.

The group velocity $v_g(k)\equiv {\rm d}\omega/{\rm d}k$ is readily
calculated from Eq.~(\ref{disp_foam}).
The relative change of $v_g(k)$ between wave numbers $k_1$ and $k_2$
is then found to be
\begin{eqnarray}
\left.\frac{\Delta c}{c}\,\right|_{\, k_1,k_2} &\equiv& 
  \left| \frac{v_g(k_1)-v_g(k_2)}{v_g(k_1)} \right| \no\\[2mm]
  &\approx& 2 \left| k_1^2 - k_2^2 \right| \, \alpha^2 \, l_{\rm foam}^2\,,
\label{theo}
\end{eqnarray}
where $\Delta c/c$ is a convenient short-hand notation and
$\alpha \approx 1/137$ the fine-structure constant.

As realized by Amelino-Camelia \emph{et al.} \cite{AEMNS98}, one can use the
lack of time dispersion in gamma-ray bursts to get an
upper bound on $\Delta c/c\,$. 
But for our purpose, it may be better to use a particular $\mathrm{TeV}$ 
gamma-ray flare of the active galaxy Markarian 421, as discussed by 
Biller \emph{et al.} \cite{B99}. Schaefer \cite{S99} obtains from this event
\begin{equation}
\left.\frac{\Delta c}{c}\:
\right|^{\rm \; Mkn\,\, 421}_{
     \begin{array}{l}
     {\scriptstyle k_1=2.5\times 10^{16} \, {\rm cm}^{-1}} \\[-2mm]
     {\scriptstyle k_2=1.0\times 10^{17} \, {\rm cm}^{-1}} 
     \end{array}}
\; < \, 2.5 \times 10^{-14} \,.
\label{exp}
\end{equation}
The reader is invited to look at Fig.\ 2 of Ref.\ \cite{B99}, 
which provides the key input for the bound (\ref{exp}), 
together with the galaxy distance $D$. In fact, the \rhs~of Eq.~(\ref{exp})
is simply the ratio of the binning interval for the gamma-ray events 
($\Delta t \approx 280\: \mathrm{s}\,$) 
over the inferred travel time ($D/c \approx 1.1\times 10^{16}\:\mathrm{s}\,$).

Combining our theoretical expression (\ref{theo}) and the astrophysical 
bound (\ref{exp}), we have the following ``experimental'' limit:
\begin{equation}
l_{\rm foam} < 1.6 \times 10^{-22} \,\,{\rm cm}\,,
\label{upperbound}
\end{equation}
where $l_{\rm foam}$ is defined by (\ref{a1b1}) and (\ref{lfoamdef}),
in terms of the autocorrelation function (\ref{correl}) for the static random 
variable $g_1(\vec x)$ from the action (\ref{Gammaphoton}). 
In the next section, we comment briefly on the possible interpretation 
of this result.

Since the spacetime foam considered in this paper
has no preferred spatial direction,
experimental limits obtained from bounds on the birefringence of
electromagnetic waves \cite{CFJ90,CK98KM02,JLMS03} do not apply. 
[See also the remarks in the penultimate paragraph of 
Sec.~\ref{sec:PhotonDispersion}.]   
Note that the background field $g_1(\vec x)$ 
vanishes on average and that the modification of the dispersion 
law in Eq.~(\ref{photon_disp_law}) is a second-order effect, 
governed by the autocorrelation of the background field.
As shown by Eq.~(\ref{disp_foam}),
the fundamental length scale $l_{\rm foam}$ appears only at 
quartic order in the photon wave number, which makes $l_{\rm foam}$
difficult to constrain (or determine) experimentally.

\section{Conclusion} 
\label{sec:conclusion}

The present article contains two main theoretical results. 
First, it has been shown in Sec.\ \ref{sec:SingleDefect} and 
Appendix \ref{sec:single_wormhole} that two particular types of  
``defects'' of a noncompact spacetime manifold generate an anomalous breaking 
of Lorentz and CPT invariance. The anomalous term of the effective gauge 
field action can be written in a simple form,  Eqs.~(\ref{anom3})
and (\ref{absf}). The spacetime defect is thus found to affect the 
photon field far away, the agent being the second-quantized vacuum of 
the chiral fermions. Inversely, the CPT anomaly can be used as a probe of 
certain spacetime structures at the very smallest scales.

Second, a modified photon dispersion law, Eq.~(\ref{photon_disp_law}),
has been found in a model which generalizes the single-defect result. 
The random background field $g_1(\vec x)$ of this model (\ref{Gammaphoton}) 
traces back to a postulated 
time-independent foam-like structure of spacetime consisting of 
many randomly oriented and randomly distributed defects.
The static random background field 
$g_1(\vec x)$ of the photon model breaks Lorentz and CPT invariance and
selects a class of preferred inertial frames.
Incidentally, the calculated photon dispersion law shows the Lorentz
noninvariance present at the microscopic scale but not the CPT violation.

This article also gives an ``experimental'' result. Following up on 
the suggestion of Refs.\ \cite{AEMNS98, B99}, it is possible to use
observations of gamma-ray bursts and $\mathrm{TeV}$  flares in active 
galactic nuclei to obtain an upper bound (\ref{upperbound}) 
on the length scale $l_{\rm foam}$ of the random background field 
$g_1(\vec x)$ of the model considered.
As such, this upper bound constitutes a nontrivial result for
a particular characteristic of a multiply connected space manifold
at the very smallest scale, granting
the relevance of the  model (\ref{Gammaphoton}) for the effects of the 
CPT anomaly (see, in particular, the discussion in
Sec.\  \ref{sec:Motivation}).

The upper bound (\ref{upperbound}) on $l_{\rm foam}$ is, of course, 
eleven orders of magnitude above the Planck length,
$l_{\rm Planck} \equiv \sqrt{G\hbar/c^3} \approx 1.6\times 10^{-33}\,\,{\rm cm}$.
But it should be realized that we have no real understanding of the
possible topologies of spacetime, be it at the very smallest scale or
the very largest. 
It is even possible that $l_{\rm foam}$ and $l_{\rm Planck}$ are
unrelated; for example, if $l_{\rm foam}$ is not a quantum effect, but
has some other, unknown, origin.

\begin{appendix}

\section{Calculations for static wormholes}
\label{sec:appendix} 
\subsection{CPT anomaly from a single wormhole}
\label{sec:single_wormhole} 

The orientable  three-space $N_3$ considered in this subsection has a single 
permanent wormhole (traversable or not). The wormhole is constructed by 
removing two identical open balls from $\bR^3$ and properly 
identifying their surfaces;
see, e.g., Sec. II A of Ref.~\cite{Fetal90} and Sec. 15.1 of Ref.~\cite{V96}
for further details. 
The removed balls have diameter $b$ and are separated by a distance
$d$ in $\bR^3$.      
The length of the wormhole ``throat'' is then zero, 
whereas the long distance between the wormhole ``mouths'' is $d$. 
As a particularly simple case, the width $b$ of the wormhole mouths
is put to zero, so that the three-space $N_3$ is essentially $\bR^3$ with two
points identified. The coordinates of these identified points are taken 
to be $\vec x = \pm (d/2)\, \hat r$, for an arbitrary unit vector $\hat r$.

Like the punctured three-manifold $M_3$ of Sec.~\ref{sec:SingleDefect}, 
this three-space $N_3$ is multiply connected and the CPT anomaly \cite{K00} 
may be expected to occur. 
In order to show this, we proceed along the same lines as in
Sec.~\ref{sec:SingleDefect}. That is, we introduce a suitable
coordinate system over the spacetime manifold $N=\bR\times N_3$
and restrict our attention to a class of background gauge fields
which allows us to trace the CPT anomaly back to the three-dimensional
parity anomaly \cite{R84, ADM85, CL89}. 

\begin{figure}
\begin{center}
\includegraphics[width=8cm]{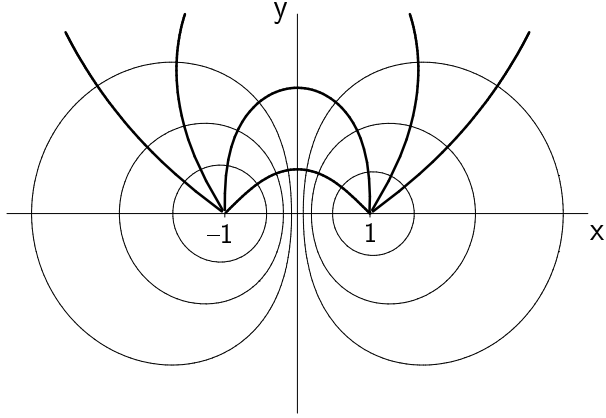}
\end{center}
\caption{Parametrization of the three-space $N_3$ 
  with a permanent wormhole at $x=\pm 1$ and $y=z=0$
  ($x$ and $y$ coordinates shown, $z$ coordinate suppressed).  
  The thin lines indicate ``equipotential''
  surfaces on which $\rho$ runs from $0$ to $\infty$ and $\phi$ from
  $0$ to $2\pi$  (shown here are $\phi=0$ and $\phi=\pi$).
  The thick lines show noncontractible loops on which $\eta$ 
  runs from $-\pi/2$ to $+\pi/2$. 
  The minimal length of these noncontractible loops  
  is given by $d$ in Eq.~(\ref{Phidef}), which has the value $2$ here.
  \label{fig:dipole}}
\end{figure}

Let $\Phi(\vec x)$ be a scalar function on $\bR^3$, so that the integral
curves of $\nabla \Phi$ are noncontractible loops on $N_3$. 
As a concrete example, take 
\begin{equation}
\Phi(\vec x) \equiv \arctan \left( \frac{d}{|\vec x-(d/2)\,\hat r|} -
    \frac{d}{|\vec x+(d/2)\,\hat r|}  \right)\,,
\label{Phidef}
\end{equation}
which resembles the potential of an electric dipole.
The manifold $N_3$ is now parametrized by the coordinates 
$\eta$, $\rho$, $\phi$, where
$\phi \in [0,2\pi)$ is the azimuthal angle around the $\hat r$-axis,
$\eta\in[-\pi/2, \pi/2)$ is defined by $\eta \equiv \Phi(\vec x)$, and
$\rho \in [0,\infty)$ is a coordinate perpendicular to $\phi$ and
$\eta$. The pair $(\rho, \phi)$ parametrizes the ``equipotential'' surfaces 
of $\Phi$. 
The equipotential surface has the topology of a two-sphere for $\eta \ne0$  
and that of a plane for $\eta=0$ ($\rho$ and $\phi$ now correspond to the 
usual polar coordinates); see Fig.\ \ref{fig:dipole}. 

For our purpose, it suffices to establish
the CPT anomaly for one particular class of gauge fields.
We, therefore, take the gauge fields to be independent of $\eta$ and to have 
no component in the direction of $\eta$. 
These gauge fields will be indicated by a double prime. 
With appropriate vierbeins, the fermionic fields are assumed to have 
periodic boundary conditions in $\eta$.

The anomaly can be calculated for the plane $\eta=0$ by
making use of the 
three-dimensional parity anomaly. Choosing $\hat r \equiv
(0,0,1)$, the relevant Chern--Simons term reads
\begin{eqnarray}  
&&\omega_{\rm CS}[B^{\prime\prime}_0, B^{\prime\prime}_1,
B^{\prime\prime}_2]
\no \\ 
&&
= \frac{1}{16\pi^2}\,\, q_\lambda^{\prime\prime}\, \,
\epsilon^{\mu\nu\kappa\lambda}\,\,  \tr \left(
  B^{\prime\prime}_{\mu\nu} B^{\prime\prime}_\kappa - \tfr{2}{3}\,
  B^{\prime\prime}_\mu  B^{\prime\prime}_\nu B^{\prime\prime}_\kappa
\right)\,,
\end{eqnarray} 
with
\begin{equation}
(q^{\prime\prime\mu}) \equiv\left(0, \,\,\vec
    e_\eta\bigr|_{\eta=0} \right) =  (0,0,0,-1)\,,
\end{equation}
where $\vec e_\eta$ denotes the unit vector in the $\eta$-direction, $\vec
e_\eta \equiv \nabla \Phi / |\nabla \Phi|$.
The three-dimensional anomaly term is then
\begin{align}
\int_{\bR^3} \dint{x^0}\dint{x^1}\dint{x^2} \pi\, s_0\, 
\omega_{\rm CS}[B^{\prime\prime}_0 , B^{\prime\prime}_1, B^{\prime\prime}_2]
 \,,
\label{whanom3d}
\end{align}
with an odd integer $s_0$  from the ultraviolet regularization
(in the following, we take $s_0= 1$). With the vector
\begin{equation}
\omega^j[B''] \equiv \frac{1}{16\pi}\, \epsilon^{\mu\nu\kappa j}\,
 \tr \left(
  B^{\prime\prime}_{\mu\nu} B^{\prime\prime}_\kappa - \tfr{2}{3}\,
  B^{\prime\prime}_\mu  B^{\prime\prime}_\nu B^{\prime\prime}_\kappa
\right) \,,
\end{equation}
the anomaly term (\ref{whanom3d}) may be written as
\begin{equation}
- \int \dint{x^0} \! \int_{\cS_0} \vec \omega[B'']\cdot\, \dint{\vec S}\,,
\label{whanomS0}
\end{equation}
where $\cS_\eta$ denotes the surface of constant $\eta$.

Since the fields considered are independent of $\eta$, the result 
(\ref{whanomS0}) is independent of the particular equipotential surface chosen. 
Hence,
\begin{equation}
-\int \dint{x^0} \! \int_{\cS_\eta} \vec \omega[B'']\cdot\, \dint{\vec S}
\end{equation}
is independent of $\eta$ and can be averaged over $\eta$.
The anomaly term can then be written as a four-dimensional
integral over the spacetime manifold $N=\bR\times N_3$,
\begin{eqnarray}
&&-\int \dint{x^0} \! \int_{\cS_0} \vec \omega[B'']\cdot\, \dint{\vec S}
\no \\[1mm]  
&&
= -\frac{1}{\pi} \int_{-\pi/2}^{\pi/2} \dint{\eta} \int \dint{x^0} \!
\int_{\cS_\eta} \vec \omega[B'']\cdot\, \dint{\vec S}
\no \\[1mm]
&&
= - \frac{1}{R[\Phi]}\, \int_N \dint{^4 x} \bigl|\nabla \Phi(\vec
x)\bigr| \, \bigl(\vec \omega[B''] \cdot \vec e_\eta \bigr)
\no \\[1mm] 
&&
= \frac{1}{R[\Phi]}\, \int_N \dint{^4 x} \Phi(\vec x)\,\nabla \cdot 
\vec\omega[B'']\,,
\end{eqnarray}
where $R[\Phi]$  is the range of values of the function $\Phi(\vec x)$, 
i.e., the length of the coordinate interval of $\eta$. For the particular 
function $\Phi$ given in Eq.\ (\ref{Phidef}), one has $R[\Phi]=\pi$.

If $B''_\mu$ contains only an electromagnetic component $A_\mu''$, the
anomalous contribution to the effective action is
\begin{equation}
\int_N \dint{^4x} \,\frac{\alpha\,\Phi(x)}{R[\Phi]} \;
\epsilon^{\kappa\lambda\mu\nu} \,F''_{\kappa\lambda}(x)\, F''_{\mu\nu}(x)\,,
\label{AbeliananomN} 
\end{equation}
up to an overall constant of order $1$, which depends on the details of the 
theory considered. Here, the field strength 
$F''_{\mu\nu}(x)$ is defined by
$\partial_\mu A''_\nu(x) - \partial_\nu A''_\mu(x)$
and $\Phi(x)$ is the scalar function (\ref{Phidef}), possibly with a 
different normalization. The general structure of the term 
(\ref{AbeliananomN}) is of the form as given by Eq.~(\ref{anom3}), 
now with a functional $f_{N}(x;A]$ in the integrand.

\subsection{Random background field and \boldmath $l_{\rm foam}$}
\label{sec:wormhole_gas}

In this subsection, we present a simple background field $g_1(\vec x)$ 
to mimic the anomalous effects from a random distribution of wormholes. 
The photon model (\ref{Gammaphoton}) incorporates, in this way,
the basic features of the CPT anomaly for a single
wormhole as found in the previous subsection of this appendix. 
For purely technical reasons, we use, instead of the function (\ref{Phidef}), 
a function $\widehat{\Phi}$ which is the direct difference of two
``monopole'' contributions, 
\begin{equation}
\widehat{\Phi}(\vec x) = K\, \left[ h\Bigl(\vec x -\frac{\delta}{2}\, \hat r \Bigr)
                         - h\Bigl(\vec x +\frac{\delta}{2}\, \hat r \Bigr)
                     \right] \,, 
\label{Phiprimedef}
\end{equation}
with a displacement parameter $\delta>0$ and a normalization factor
$K>0$ (see below). In addition, the profile function $h(\vec x)$ is assumed 
to be isotropic, monotonic, finite, and integrable.
The random background field is a superposition of these ``dipole''
potentials, with random locations ($\vec x_n$) and directions 
($\hat r_n$). The random background field also includes a factor
$\alpha\,$; cf. Eq.~(\ref{AbeliananomN}). 

For $N \equiv (4/3)\pi R^3/a^3$ dipoles in a ball of radius $R$ 
(with an average separation $a$ between the different wormholes), we have 
\begin{eqnarray}
g_N(\vec x) &=& \alpha\,K\, \sum_{n=1}^N \Bigl[
     h\Bigl(\vec x-\vec x_n -\frac{\delta}{2}\, \hat r_n \Bigr)
\no \\  
&&
   - h\Bigl(\vec x-\vec x_n +\frac{\delta}{2}\, \hat r_n \Bigr)
  \Bigr]\,,
\end{eqnarray}  
where $\vec x_n$ are random vectors with $|\vec x_n|<R$ and $\hat
r_n$ randomly chosen unit vectors.
The final background field for the photon model (\ref{Gammaphoton})
is given by the infinite volume limit,
$g_1(\vec x) = \lim_{N\to \infty} g_N(\vec x)$. 
The relevant autocorrelation function is 
\begin{align}
C_1(\vec y) &\equiv \lim_{N\to \infty} \frac{1}{(4/3)\pi R^3}\int_{|
\vec x|<R} \dint{^3 \vec x} g_N(\vec x) \, g_N(\vec x+\vec y)\no \\[1ex]
&= \alpha^2\,\frac{K^2}{a^3} \,\, \text{av}_{\hat r} \int  \dint{^3 \vec x}  
\left[ h \Bigl(\vec x -\frac{\delta}{2}\, \hat r \Bigr)
     - h \Bigl(\vec x +\frac{\delta}{2}\, \hat r \Bigr)  \right] \nonumber \\
&\quad \times 
\left[ h \Bigl(\vec x+\vec y -\frac{\delta}{2}\, \hat r \Bigr)
     - h \Bigl(\vec x+\vec y +\frac{\delta}{2}\, \hat r \Bigr) \right] \,,
\end{align}
where $\text{av}_{\hat r}$ denotes the average over the orientations
$\hat r$.

The typical length scale which enters the photon dispersion law 
(\ref{photon_disp_law}) is defined by Eqs.\
(\ref{a1b1}) and (\ref{lfoamdef}), 
\begin{align}
l_{\rm foam}^2 &\equiv
 \frac{1}{30\alpha^2} \int \dint{^3\vec y} \frac{C_1(|\vec y|)}{|\vec
  y|}\,.
\end{align}
After a suitable shift of integration variables, one finds
\begin{equation}
l_{\rm foam}^2 = \frac{1}{15} \left(A-B\right)\,,
\label{lfoamAB}
\end{equation}
in terms of two integrals,
\begin{subequations} 
\begin{eqnarray}
A &\equiv&  K^2\, \int \dint{^3\vec y}\dint{^3\vec x}\, \frac{1}{|\vec y|}\,
\frac{1}{a^3} \,  h\left(\vec x \right)  h\left(\vec
  x+\vec y \right)\,,\\
B &\equiv&  K^2\, \int \dint{^3\vec y}\dint{^3\vec x}\, \frac{1}{|\vec
  y-\hat r\,\delta|}\,\frac{1}{a^3} \, 
 h\left(\vec x \right)  h\left(\vec
  x+\vec y \right)\,.
\end{eqnarray}
\end{subequations} 
The integral $B$ is, in fact, independent of the direction $\hat r$
and there is no need to average over it.
By using spherical coordinates  $(y, \theta_y, \phi_y)$
with respect to the $\hat r$-axis for $\vec y$ and $(x,\theta_x, \phi_x)$ 
around the $\vec y$-axis for $\vec x$, one finds after performing 
three angular integrals:   
\begin{eqnarray}  
B &=& A+  K^2\, \frac{2\,(2\pi)^2}{a^3} 
\int_0^\delta \!\dint{y}\! \!\int_0^\infty \!\!\dint{x} \!\int_0^\pi
\!\dint{\theta_x}  
\sin\theta_x\; x^2y^2\,
\no \\[1ex] 
&&
\times \, \left(\frac{1}{\delta}  - \frac{1}{y} \right) 
h(\vec x)\, h(\vec x + \vec y)\,.
\end{eqnarray}  
With Eq.\ (\ref{lfoamAB}), this gives
\begin{eqnarray}  
l_{\rm foam}^2 &=& \frac{K^2}{15\pi\, a^3} \int_{|\vec y|\le \delta}
\dint{^3 \vec y} \left(\frac{1}{|\vec y|} -\frac{1}{\delta} \right) 
\no \\[1ex] 
&& \times \, \int \dint{^3\vec x} \, h(\vec x) h(\vec x+ \vec y) \,.
\end{eqnarray}

\begin{figure}
\begin{center}
\includegraphics[height=6cm]{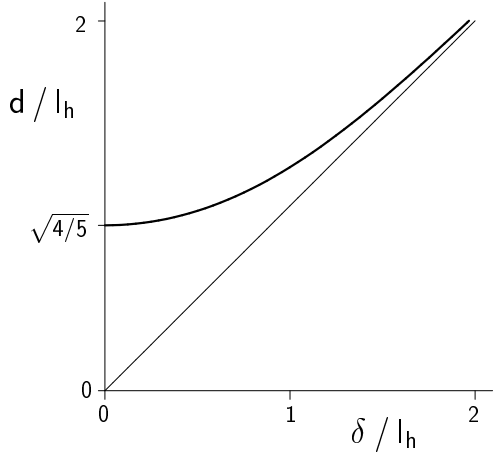}
\end{center}
\caption{Long distance $d$ between the wormhole mouths vs.\  parameter
         $\delta$ entering the profile functions $h$ of Eqs.\
         (\ref{Phiprimedef}) and (\ref{profx4}),
         both in units of the width $l_h$. \label{fig:deltad}}
\end{figure}

In order to get a concrete result for $l_{\rm foam}$, we choose
the following profile function:
\begin{equation}
h(\vec x) =\left(\frac{l_h^2}{|\vec x|^2+l_h^2}\right)^2\,, 
\label{profx4}
\end{equation}
with a scale parameter $l_h>0$. It is important to note that the peaks of
$\widehat{\Phi}(\vec x)$ from Eq.\ (\ref{Phiprimedef})
are not located at $\pm (\delta/2)\, \hat r$, but rather at
$\pm (d/2)\, \hat r$ for an effective distance $d=d(\delta,l_h)$;
see Fig.\ \ref{fig:deltad}. 
The normalization factor $K=K(\delta,l_h)$ in Eq.\ (\ref{Phiprimedef})
is chosen such that  
$\widehat{\Phi}(\vec x)$ has values $\pm 1$ at these peaks. 
The expressions for $d$ and $K$ are somewhat involved and need not 
be given explicitly.

\begin{figure}
\begin{center}
\includegraphics[height=6cm]{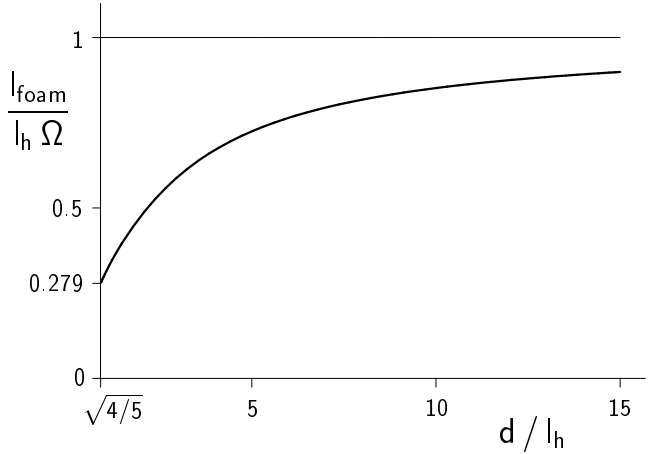}
\end{center}
\caption{Length scale $l_{\rm foam}$ entering the photon dispersion law 
         (\ref{disp_foam}) vs.\ individual wormhole scale 
         $d\,$, with the further definition  
         $\Omega^2 \equiv (\pi^2/15)\, l_h^3/a^3$.
\label{fig:lfoam} }
\end{figure}

For the profile function (\ref{profx4}), we finally obtain 
\begin{equation}
l_{\rm foam}^2 = \frac{\pi^2}{15}\, \frac{l_h^5}{a^3} \; K(\delta, l_h)^2\,
\left( 1-  \frac{2l_h}{\delta} \arctan \frac{\delta}{2l_h}\right)\,,
\end{equation}
with average separation $a$ between the different wormholes and parameters
$\delta$ and $l_h$ for the individual wormholes; cf. Eqs.~(\ref{Phiprimedef}) 
and (\ref{profx4}). As mentioned above, precisely the quantity 
$l_{\rm foam}^2$ enters the quartic term of the photon dispersion 
law (\ref{disp_foam}). For $\delta/l_h \to \infty$, one finds 
\begin{equation}
l_{\rm foam}^2\Bigr|_{\delta/l_h \to \infty}    
= \frac{\pi^2}{15}\, \frac{l_h^5}{a^3}\,, 
\end{equation}
which is essentially the same result as for a random distribution of
``monopoles'' 
with profile function $h(\vec x)$.
For $\delta/l_h \downarrow 0$, on the other hand, the effective
distance $d$ approaches the value $\sqrt{4/5}\,\, l_h$ 
and  $l_{\rm foam}^2$ becomes
\begin{equation}
l_{\rm foam}^2\Bigr|_{\delta/l_h=0} \approx  
\frac{\pi^2}{15}\, \frac{l_h^5}{a^3}
\, \times \, (0.279)^2 \,.
\end{equation}
Figure \ref{fig:lfoam} shows the rather weak dependence of $l_{\rm foam}$
on the individual wormhole scale $d$.

\end{appendix}

\end{document}